\documentclass[12pt]{article}

\newcommand{\x}{arXiv:}
\newcommand{\m}{\mathrm}

\usepackage{graphicx}

\begin{document}
\thispagestyle{empty}
\begin{center}

\null \vskip-1truecm \vskip2truecm

{\Large{\bf \textsf{Fragile Black Holes and an Angular Momentum Cutoff in Peripheral Heavy Ion Collisions}}}

{\large{\bf \textsf{}}}

{\large{\bf \textsf{}}}

\vskip1truecm

{\large \textsf{Brett McInnes
}}

\vskip1truecm

\textsf{\\ Centro de Estudios Cient}$\acute{\textsf{i}}$\textsf{ficos (CECs), Valdivia, Chile}
\vskip0.1truecm
\textsf{\\ and}
\vskip0.1truecm
\textsf{\\ National
  University of Singapore}\footnote{Permanent Address}
  \vskip0.3truecm
\textsf{email: matmcinn@nus.edu.sg}\\

\end{center}
\vskip1truecm \centerline{\textsf{ABSTRACT}} \baselineskip=15pt

\medskip
In collisions of heavy ions at extremely high energies, it is possible for a significant quantity of angular momentum to be deposited into the Quark-Gluon Plasma which is thought to be produced. We develop a simple geometric model of such a system, and show that it is dual, in the AdS/CFT sense, to a rotating AdS black hole with a topologically planar event horizon. However, when this black hole is embedded in string theory, it proves to be unstable, for all non-zero angular momenta, to a certain non-perturbative effect: the familiar planar black hole, as used in most AdS/CFT analyses of QGP physics, is ``fragile". The upshot is that the AdS/CFT duality apparently predicts that the QGP should always become unstable when it is produced in peripheral collisions. However, we argue that holography indicates that relatively low angular momenta \emph{delay} the development of the instability, so that in practice it may be observable only for peripheral collisions involving favourable impact parameters, generating extremely large angular momenta. In principle, the result may be holographic prediction of a cutoff for the observable angular momenta of the QGP, or perhaps of an analogous phenomenon in condensed matter physics.

\newpage

\addtocounter{section}{1}
\section* {\large{\textsf{1. The Spinning Quark-Gluon Plasma}}}
The RHIC \cite{kn:phobos} and LHC \cite{kn:schuk}\cite{kn:schuk2} experiments study ultra-high energy collisions of heavy nuclei. Such collisions can be either \emph{central} or \emph{peripheral}, meaning in the latter case that the impact parameter is not negligible. In the central case, collisions at the highest energies are generally agreed to produce a \emph{quark-gluon plasma} [QGP], with properties which can be studied in a variety of ways, including the use of the AdS/CFT correspondence [see for example \cite{kn:solana}\cite{kn:pedraza}].
In this approach, the QGP is studied by means of a dual system built around a thermal asymptotically AdS black hole.

Peripheral collisions present a number of novel aspects. In particular, it has been argued \cite{kn:liang}\cite{kn:bec} [see also \cite{kn:huang} for recent developments] that, for the most energetic peripheral collisions, such as those at the LHC, a significant proportion of the angular momentum is deposited directly into the interaction zone, that is, into the QGP itself. It can be shown that this may result in a number of striking and characteristic phenomena, which should allow experimental confirmation that the QGP in these collisions does indeed carry a large angular momentum; in particular, one can hope to observe quark polarization effects\footnote{The corresponding anisotropy, enforced dynamically by angular momentum conservation, is not to be confused with the \emph{initial} anisotropy which obtains in the immediate aftermath of the collision, and which decays extremely quickly: see for example \cite{kn:tranc}. For the relationship of this particular anisotropy with the elliptic flow, see the discussion in \cite{kn:schalm}; for the holography of the elliptic flow itself, see \cite{kn:shin}.}.

It may seem obvious that one should use some kind of AdS-Kerr black hole to construct a holographic model of this situation, but caution is called for here. One cannot, of course, obtain the Kerr metric simply by rewriting the Schwarzschild metric in rotating coordinates: angular momentum is a ``charge" for black holes, as fundamental as electric charge or indeed mass. This raises the possibility that a holographic description of the rotating QGP may well predict unexpected dynamical effects, and not merely reproduce the kinematic effects predicted in \cite{kn:liang}\cite{kn:bec}. We shall argue that this is precisely what happens.

Atmaja and Schalm \cite{kn:schalm} have in fact used AdS-Kerr black holes to construct a very interesting AdS/CFT model of the angular momentum contribution to the overall anisotropy and its consequences for jet quenching. [AdS-Kerr black holes have also been used to construct holographic models of rotating superconductors \cite{kn:sonner}.] They used the standard AdS-Kerr solution [in four dimensions, so that the field theory is defined on a three-dimensional boundary spacetime] with a topologically spherical event horizon \cite{kn:carter}\cite{kn:hawrot}, and \emph{not} [a rotating version of] the ``planar" black hole which is conventionally employed in applications of AdS/CFT to heavy ion physics. While the results were remarkably good in this specific application, topologically spherical event horizons have several serious drawbacks from the point of view of further applications to peripheral collisions, as we now discuss.

The most obvious problem is simply that the geometry of the spatial sections of the boundary does not describe the actual geometry of a peripheral heavy ion collision. As is explained in \cite{kn:liang}\cite{kn:bec} [see also \cite{kn:gao}], the fraction\footnote{In fact, most of the angular momentum is carried away by the parts of the nuclei that do not overlap; the fraction transferred to the QGP is significant only for extremely high-energy collisions. In particular, this effect is expected to be markedly more important at the LHC than at the RHIC.} of the initial angular momentum that is deposited into the QGP arises from the non-uniform distribution of the constituent nucleons in the direction transverse to the direction of motion of the nuclei. This leads to a \emph{particular distribution of momentum} in the aftermath of the collision,
and it is this crucial feature that one wishes to reproduce in a geometric model of the spinning QGP. One can describe the situation in a simplified way as follows.

\begin{figure}[!h]
\centering
\includegraphics[width=1.5\textwidth]{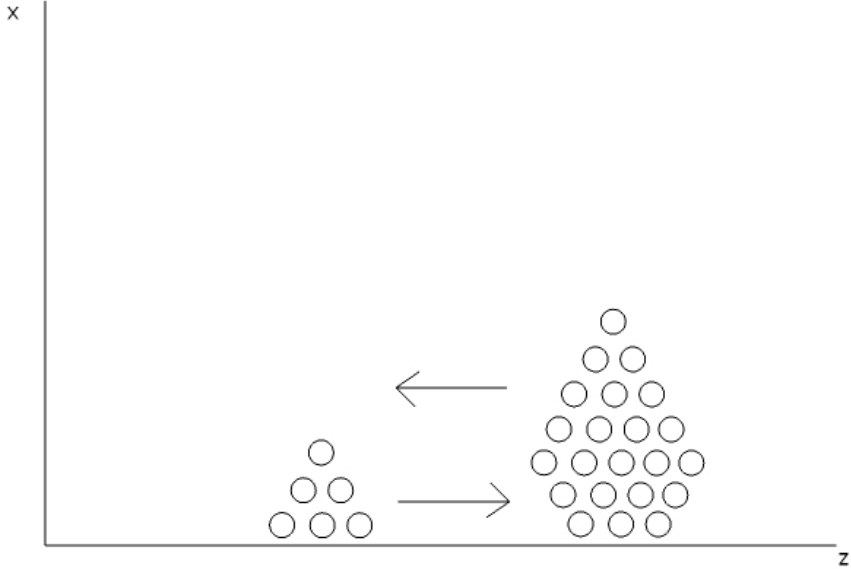}
\caption{Model of Peripheral Collision, Showing Transverse Non-Uniformity.}
\end{figure}
\begin{figure}[!h]
\centering
\includegraphics[width=1.4\textwidth]{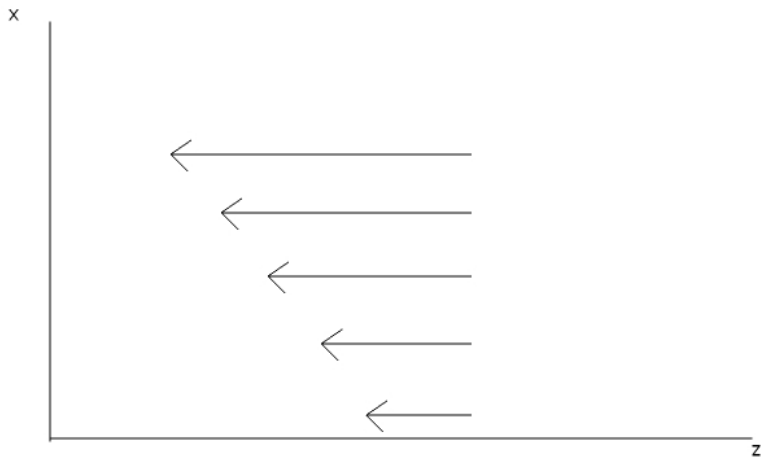}
\caption{The Post-Collision Momentum Distribution.}
\end{figure}

Figure 1 [which is adapted from \cite{kn:bec}] is a schematic representation of the peripheral collision of two identical nuclei, portrayed as diamond-shaped collections of nucleons. [For convenience we have removed the parts of the nuclei below the horizontal axis: the region of interest is the one immediately above that axis.] The point of the diagram is to show that, due to the transverse non-uniformity, the nucleons in one nucleus located at a given distance from the axis will \emph{not} in general meet the same number of nucleons moving in the opposite direction.
This gives rise to a shearing motion within the QGP, shown schematically in Figure 2 [also from \cite{kn:bec}].

The extent of this effect depends on the impact parameter; we are mainly interested in the case in which the impact parameter is such that the proportion of angular momentum deposited into the QGP is maximal [about 29$\%$ in the situation discussed in \cite{kn:bec}]. This generates a pattern such that there is zero net momentum along a certain neutral axis [the z axis in Figure 2], but with increasing momenta away from that axis until the edge of the overlap region is reached, so that Figure 2 should be interpreted as being effectively \emph{finite} in the x direction; this also avoids arbitrarily large momenta. [In \cite{kn:bec},\cite{kn:schalm}, and here, the y-axis, perpendicular to the plane of the diagrams and parallel to the angular momentum vector, plays only a minor role; each y = constant slice is treated independently.]

This is quite unlike the geometry of the spatial sections at infinity of the usual [four-dimensional] topologically spherical AdS-Kerr metric. These have the topology of the two-dimensional sphere, endowed with a specific [non-round] geometry. Compare this with Figure 2. The sphere rotates differentially [that is, at an angular velocity which depends on the azimuthal angle], which does reflect one aspect of what one sees in Figure 2, but the angular velocity vanishes only at a pair of points, not along a neutral axis. The problem is simply that, for this application, we do not want either a curved geometry or spherical topology for the spatial sections at infinity: in order to generate a geometric model of Figure 2, we want a \emph{flat} geometry in the x-z plane, and a position-dependent distortion which vanishes along some axis but steadily increases away from that axis.

A much less obvious, but very important drawback of using the topologically spherical AdS-Kerr black hole [in this application] is seen when the string physics in the bulk is examined more closely. In particular, any distortion of the bulk geometry will have a profound effect on the areas and volumes of extended objects like branes, and Seiberg and Witten \cite{kn:seiberg} showed that this can give rise to a serious instability in that sector of the theory. [Essentially, the geometric distortion affects the brane action in such a manner that the result is a pair-production instability for branes.] In \cite{kn:74} it was shown that, despite the fact that the AdS-Kerr geometry often differs very greatly from that of an AdS-Schwarzschild black hole, this effect is never a problem even for rapidly rotating topologically spherical AdS black holes. This result is desirable in the application to astrophysical black holes, which do of course have topologically spherical event horizons, but it is \emph{not} welcome here. For it means that, by using an inappropriate geometry for the event horizon [and thus at infinity], we run the risk of obscuring an instability \emph{which might actually be present} but which would only appear in the theory if we were able to use a planar black hole to reproduce a structure more similar to that portrayed in Figure 2.

In the following section we shall develop a simple geometric model, based on a differentially rotating [or shearing] \emph{flat} two-dimensional space, which does correctly represent the situation shown in Figure 2. The next step is to find the AdS/CFT dual; this clearly must be a rotating four-dimensional black hole with an event horizon having planar topology. An example of such a black hole was discovered by Klemm, Moretti, and Vanzo \cite{kn:klemm}, and we show that their solution does indeed induce at infinity a special case of the geometry described in Section 2. We are therefore in a position to investigate whether the Seiberg-Witten instability arises here, in the absence of the protective effect of a topologically spherical event horizon. We show in Section 4 that it does, for all non-zero angular momenta: non-rotating planar AdS black holes are themselves stable, but they are \emph{fragile} \cite{kn:fragile} in this sense. The interpretation of this result is the subject of Section 5. We show that, for low angular momenta, simple classical causality in the bulk [interpreted holographically as setting a time scale for the dual field theory] may prevent the instability from developing quickly enough to affect the system in the limited time available before hadronization occurs. Extremely high angular momenta modify the bulk geometry in such a way that this delaying effect is reduced, however, and this may mean that holography reveals a cutoff on the possible angular momenta observable in peripheral heavy ion collisions, or perhaps in analogous condensed-matter systems.

\addtocounter{section}{1}
\section* {\large{\textsf{2. A Geometric Model of Peripheral Collisions}}}
The topologically spherical AdS-Kerr metric, expressed in terms of generalized Boyer-Lindquist coordinates, induces at infinity [see \cite{kn:74} for a discussion] a metric of the form
\begin{equation}\label{SPHERE}
\m{g(TSAdSK)_{\infty}\;=\;-\,d\tau^2 \;-\;{2\,(a/L)\,sin^2(\theta)\,d\tau d\phi \over 1 - (a/L)^2} \;+\; {d\theta^2 \over 1 - (a/L)^2cos^2(\theta)} \;+\; {sin^2(\theta)d\phi^2 \over 1 - (a/L)^2} },
\end{equation}
where $\tau$ is a [dimensionless] time coordinate, $\theta$, $\phi$ are the usual spherical angles, a is the ratio of the ADM angular momentum to the ADM mass,
and L is the asymptotic curvature radius. This is the geometry used by Atmaja and Schalm \cite{kn:schalm} to model the spinning plasma; we hope to modify it in order to produce a more accurate representation of the system pictured in Figure 2. Notice in particular that the off-diagonal components induce a shearing motion, parameterized by the angular momentum, in the two-sphere parameterized by $\theta$ and $\phi$, since they depend non-trivially on both $\theta$ and the parameter a.

Consider a flat two-dimensional space with Cartesian coordinates [see Figure 2] z and x, with x being vertical. Regard this space as a model for the spatial sections of a three-dimensional ``peripheral collision spacetime" with metric
\begin{equation}\label{A}
\m{g_{PC} \;=\; - \, dt^2 \;-\; 2\Upsilon(x \,,\alpha)^2 dtdz \;+\; dx^2 \;+\; dz^2. }
\end{equation}
This represents a flat plane which is being sheared with respect to the z-axis; the latter remains stationary, provided that the function $\m{\Upsilon(x \,,\alpha)}$ vanishes for x = 0. Clearly we want $\m{\Upsilon(x \,,\alpha)}$ to be a [dimensionless] increasing function of x. The rate of the shearing is determined by the dimensionless positive constant $\alpha$; again, $\m{\Upsilon(x \,,\alpha)}$ should be an increasing function of this parameter.

We propose this as a simple geometric model of the situation portrayed in Figure 2: that is, the plasma is taken to be co-moving with respect to the given coordinates\footnote{The reader should bear in mind that, in the application to heavy ion physics, the AdS/CFT duality involves a \emph{non-gravitational} theory on the boundary. Thus, the geometry defined by g$_{\m{PC}}$ should simply be interpreted as a fixed background; there is no gravitational field equation on the boundary.}, and so it takes part in the shearing of the underlying geometry. In this application, the constant $\alpha$ should be determined by the density of the angular momentum and energy transferred to the interaction zone, that is, by the initial conditions of the collision; in fact, it proves to be just the ratio of these two densities. [Because we treat each y = constant slice separately, we assume that the ratio of the three-dimensional angular momentum density to the energy density ---$\,$ the only observational parameter we shall need apart from the temperature ---$\,$ can be modeled by the ratio of the corresponding two-dimensional densities. Note that the individual densities are not constant in time, but we shall assume that their ratio is constant.] On the other hand, the form of $\m{\Upsilon(x \,,\alpha)}$ as a function of x is in principle determined by the ``thickness function" \cite{kn:bec}, the integral of the nucleon density with respect to z; it provides a description of the transverse distribution of nucleons in the colliding nuclei.

Because the pattern shown in Figure 2 is abruptly cut off at the edge of the overlap region, values of x beyond a certain value [fixed by the size of the nuclei and the impact parameter of the collision] are not meaningful here [and, in any case, causality in the original physical system forbids an arbitrary extension of the pattern in the x direction]. Similarly, the interaction zone is of course finite also in the z direction. Thus the model is really defined on a finite [but \emph{topologically trivial}] closed rectangle which lies along the z axis: 0 $\leq$ x $\leq$ X, 0 $\leq$ x $\leq$ Z, for some physically determined values of X and Z.

It will be convenient to introduce a constant L with the dimension of length; this specifies the overall scale in Figure 2. We then define new dimensionless coordinates by
\begin{equation}\label{B}
\m{t \;=\;L\tau, \;\;\;z \;=\;L\zeta, \;\;\; x \;=\;L\xi, }
\end{equation}
where the ranges of $\tau$, $\zeta$, and $\xi$ are in principle infinite [but those of $\xi$ and $\zeta$ are in practice from zero to finite values X/L, Z/L, as explained above]. The metric is then
\begin{equation}\label{C}
\m{g_{PC} \;=\; L^2\Big[- \, d\tau^2 \;-\; 2\Upsilon(\xi \,,\alpha)^2 d\tau d\zeta \;+\; d\xi^2 \;+\; d\zeta^2\Big]. }
\end{equation}
In order to obtain a holographic description of this situation, we now need to find an asymptotically AdS black hole which induces this geometry at infinity.

\addtocounter{section}{1}
\section* {\large{\textsf{3. A Dual Black Hole}}}
Because Anti-de Sitter spacetime does not satisfy the dominant energy condition, it permits \cite{kn:greg1}\cite{kn:greg2}\cite{kn:greg3} the existence of black holes having non-spherical topology for the event horizon, and such holes have been constructed explicitly \cite{kn:lemmo}; in particular, there are asymptotically AdS black holes with topologically \emph{planar} event horizons. These ``planar black holes" are precisely the ones which appear in applications of the AdS/CFT correspondence to heavy ion physics.

Klemm, Moretti, and Vanzo \cite{kn:klemm} discovered spacetime metrics corresponding to four-dimensional, ``rotating", asymptotically AdS black holes with topologically planar event horizons. However, as the event horizons of these objects are not compact [and cannot be ---$\,$ see the erratum for \cite{kn:klemm}], it is not immediately clear what ``rotating" means here. Furthermore, for the same reason, the mass of such an object is formally infinite. Before discussing the details of the KMV metrics, let us try to clarify these points.

We begin with \emph{non-rotating} (n+2)-dimensional AdS-Schwarzschild black holes having \emph{compact} event horizons of constant curvature k = $\{ -1, 0, +1 \}$ [so that, when k = 0, we have event horizons with the topology of a torus or a non-singular quotient of a torus]. If the asymptotic AdS curvature is $-1$/L$^2$, the metric is
\begin{eqnarray}\label{DACE}
\m{g(AdSSch^k_{n+2})} = -\, \m{\Bigg[{r^2\over L^2}\;+\;k\;-\;{16\pi M\over n V[W^k_n] r^{n-1}}\Bigg]dt^2\;} + \m{\;{dr^2\over {r^2\over L^2}\;+\;k\;-\;{16\pi M\over n V[W^k_n] r^{n-1}}} \;+\; r^2\,d\Omega^2[W_n^k].}
\end{eqnarray}
Here $\m{d\Omega^2[W_n^k]}$ is the n-dimensional ``angular" part of the metric; it is a metric of constant curvature k on an n-dimensional space W$\m{^k_n}$ with [dimensionless] volume $\m{V[W^k_n]}$. [Thus, for example, for the unit two-sphere  $\m{V[W^1_2] = 4\pi}$; for a cubic 2-torus with angular coordinates having a common periodicity 2$\pi$K ---$\,$ note that this is just one possible shape for a topologically toral space ---$\,$ we have $\m{V[W^0_2] = 4\pi^2K^2}$; and so on.] The constant M is then precisely the physical [ADM] mass.

Let us consider the k = 0 case, and set n = 2. Here the event horizon has been compactified, but it can be ``decompactified" as follows. Define M$^*$ = M/V[W$^0_2$]. Then the [two-dimensional] energy density of the black hole, evaluated [for example] at the event horizon, where r = r$_{\m{h}}$, is just M$^*$/r$^2_{\m{h}}$. Now let both M and V[W$^0_2$] tend to infinity, in such a way that M$^*$ remains finite. Then we have a planar black hole, but with the metric determined by the \emph{finite} quantity M$^*$:
\begin{eqnarray}\label{DALMATIA}
\m{g(AdSSch^P_{2})} = -\, \m{\Bigg[{r^2\over L^2}\;-\;{8\pi M^*\over r}\Bigg]dt^2\;} + \m{\;{dr^2\over {r^2\over L^2}\;-\;{8\pi M^*\over r}} \;+\;r^2\Big[d\psi^2\;+\;d\zeta^2\Big]},
\end{eqnarray}
where $\psi$ and $\zeta$ are dimensionless coordinates on the plane. We see that r$_{\m{h}}$ is determined only by M$^*$, and so it too remains finite; the event horizon retains all of its usual local properties ---$\,$ that is, it is still classically a one-way membrane, it still has a Hawking temperature, and so on, which is why we continue to refer to it as a [planar] black hole. The two-dimensional energy density on the horizon is still given by M$^*$/r$^2_{\m{h}}$. Thus M$^*$, while \emph{not} the mass of the black hole, plays a similar role.

We now imagine that the planes r = constant in this geometry are subjected to a shearing motion like the one portrayed in Figure 2. Then the system acquires an infinite amount of angular momentum, just as it has an infinite mass. However, we can define a quantity J$^*$ analogous to M$^*$: that is, J$^*$/r$^2_{\m{h}}$
is the two-dimensional angular momentum density of the black hole, evaluated at the event horizon\footnote{``Angular momentum" here means ADM angular momentum. The fact that J$^*$ computes the ADM angular momentum of the KMV black hole discussed below follows from the calculations in \cite{kn:klemm}, suitably adapted from the compactified case studied there to the planar case we need here.}. It is in this sense that a planar black hole may be said to ``rotate". [We evaluate the energy and angular momentum densities at the horizon, simply because this is a surface distinguished by the geometry of the black hole itself; note however that we are really only interested in the \emph{ratio} of these densities, which we hope to interpret holographically as the ratio of the analogous densities in the field theory, just as the Hawking temperature of the hole is to be interpreted as the temperature of the field theory.]

We may now turn to the KMV black holes. The spacetime geometry at infinity of these objects can have spatial sections which are either flat or negatively curved. Here we are only concerned with the former case, and we shall refer to these objects as the KMV$_0$ black holes. These black holes are electrically and magnetically neutral; for applications to the QGP one really needs the generalization to the charged case, but the neutral case is sufficient to establish the main point of the present work.

We will write the metric in a form that manifestly reduces to the non-rotating planar black hole metric (\ref{DALMATIA}) given above when the angular momentum parameter J$^*$ vanishes. Klemm et al. \cite{kn:klemm} use a parameter a which they show is the ratio of the ADM angular momentum to the ADM mass in the compact case. In the planar case the corresponding quantity is  a = J$^*$/M$^*$, which can be interpreted, as above, as the ratio of the angular momentum and energy densities computed at the event horizon. Using coordinates (t, r, $\psi$, $\zeta$) as above, one finds that the metric takes the form
\begin{equation}\label{L}
\m{g(KMV_0) = - {\Delta_r\Delta_{\psi}\rho^2\over \Sigma^2}\,dt^2\;+\;{\rho^2 \over \Delta_r}dr^2\;+\;{\rho^2 \over \Delta_{\psi}}d\psi^2 \;+\;{\Sigma^2 \over \rho^2}\Bigg[\omega\,dt \; - \;d\zeta\Bigg]^2},
\end{equation}
where, as before, the asymptotic curvature is $-1$/L$^2$, and
\begin{eqnarray}\label{eq:M}
\rho^2& = & \m{r^2\;+\;a^2\psi^2} \nonumber\\
\m{\Delta_r} & = & \m{a^2+ {r^4\over L^2} - 8\pi M^* r}\nonumber\\
\Delta_{\psi}& = & \m{1 +{a^2 \psi^4\over L^2}}\nonumber\\
\Sigma^2 & = & \m{r^4\Delta_{\psi} - a^2\psi^4\Delta_r = r^4 + a^2\psi^4\Big(8\pi M^*r - a^2\Big)}\nonumber\\
\omega & = & \m{{\Delta_r\psi^2\,+\,r^2\Delta_{\psi}\over \Sigma^2}\,a}.
\end{eqnarray}
Notice that this metric does indeed reduce precisely to the metric g(AdSSch$^{\m{P}}_2$), given in equation (\ref{DALMATIA}), when a = 0; it is distinguished by the fact that $\partial_{\zeta}$ continues to be a Killing vector field, though of course $\partial_{\psi}$ does not.

The physical interpretation of this geometry has a number of unfamiliar aspects, which we now discuss.

The event horizon is located at r = r$_{\m{h}}$, which is the largest root of $\Delta_{\m{r}}$. From the second expression for $\Sigma^2$ we see that the latter is an increasing function of r for positive r; since the first expression for it shows that it is positive at r = r$_{\m{h}}$, it is positive everywhere outside the black hole and so there are no difficulties arising from its vanishing. For the details of the structure inside the event horizon, and for the Penrose diagram, we again refer the reader to \cite{kn:klemm}.

The black hole is ``rotating" in the $\zeta$ direction, which however does not have a circular topology\footnote{In \cite{kn:klemm}, both planar coordinates are compactified to circles, but this will give rise to a discontinuity [see the erratum to that work] in the metric unless the time coordinate is also involved in the compactification, which would lead to closed timelike worldlines. I am grateful to Dr Marco Astorino for drawing my attention to this.}; the quantity $\omega$ is the ``angular velocity" of the black hole, which, in general, depends on both r and $\psi$. For reasons discussed in more detail below, we think of $\psi$ and $\zeta$ as taking their values on closed [finite] intervals; however, there are no topological identifications here.

The value of r at the event horizon, r$_{\m{h}}$, is related to a and M$^*$:
\begin{equation}\label{FLABBY}
\m{a^2 \; - \; 8\pi M^*r_{h} \; + \;r_{h}^4/L^2 \; = \; 0.}
\end{equation}
The black hole temperature is given [see \cite{kn:klemm}] by
\begin{equation}\label{FROGLIKE}
\m{T\;=\;{2r_{h}^3\;-\; 4\pi M^*L^2 \over 2\pi L^2r_{h}^2}.}
\end{equation}

It will be convenient to combine (\ref{FLABBY}) with (\ref{FROGLIKE}) to obtain a pair of equivalent equations:
\begin{equation}\label{FAUCET}
\m{4\pi T \;=\;{3r_{h} \over L^2} \;-\; {a^2 \over r_{h}^3};}
\end{equation}
\begin{equation}\label{FLOPPY}
\m{8\pi M^*L^2\;=\;r_{h}^3 \;+\; {a^2L^2 \over r_{h}}.}
\end{equation}
The temperature of the black hole cannot be negative: this is a way of stating cosmic censorship. Fixing a$^2$ and regarding T as a function of r$_{\m{h}}$, we see that T is an increasing function, so the requirement that T $\geq$ 0 imposes a lower bound on r$_{\m{h}}$, namely r$_{\m{h}}^4 \,\geq \,$ a$^2$L$^2$/3. This is precisely the value at which M$^*$, regarded through (\ref{FLOPPY}) as a function of r$_{\m{h}}$ [again with a$^2$ fixed], attains its minimum, so M$^*$ is always an increasing function of r$_{\m{h}}$ on the physical domain. The fact that M$^*$ must always be at least as large as its value at this minimum implies that
\begin{equation}\label{eq:N}
\m{a^2/L^2 \; \leq \; 3\times (2\pi M^*/L)^{4/3}};
\end{equation}
this is the explicit expression of cosmic censorship for these black holes.

Equality in the relation (\ref{eq:N}) defines extremality here; that is, we have equality only when the temperature vanishes. In fact, in the application to heavy ion collisions, these black holes are dual to a deconfined plasma, which is extremely hot and actually \emph{cannot} have an arbitrarily low temperature\footnote{These particular black holes are dual to a plasma at zero chemical potential $\mu$, but the point is valid at all values of $\mu$: the temperature cannot be arbitrarily small. See \cite{kn:AdSRN}\cite{kn:triple}\cite{kn:73}.}. Thus the inequality is not near to being saturated here; although the dual system may have a very large angular momentum density, the black hole is rotating rather ``slowly" in the sense that it is \emph{not} close to being extremal\footnote{In particular, then, we do not expect superradiant instabilities [see for example \cite{kn:dias}] to be important in this case.}.

In view of this observation, for our purposes we can slightly refine the inequality (\ref{eq:N}) to a \emph{strict} inequality. One can state this version in terms of dimensionless quantities, as follows:
\begin{equation}\label{eq:NOW}
\m{{1 \over 3^{1/4}}\Bigg({L \over 2\pi M^*}\Bigg)^{1/3} \; < \;\; \sqrt{{L \over a}}\,, \;\;\;\; T \; > \;0. }
\end{equation}
The reason for formulating the inequality in this particular way will become clear in the next section.

Notice that, with a$^2$ fixed, \emph{both} T and M$^*$ are [in (\ref{FAUCET}) and (\ref{FLOPPY})] increasing functions of r$_{\m{h}}$ on the physical domain, which means that the Hawking temperature is always an \emph{increasing} function of the mass parameter; all KMV$_0$ black holes behave like ``large" spherical AdS black holes in this sense. [That is, these black holes can be both arbitrarily massive and arbitrarily ``hot". For a discussion of this in the spherical case, see \cite{kn:larus}.] In other words, it appears that these objects always have a positive specific heat and so can reach equilibrium with their own Hawking radiation before they evaporate completely.

To prove this, one would however have to deal with the fact that a$^2$ would not really remain constant during the Hawking evaporation process. Normally one expects \cite{kn:page} the angular momentum to decay more rapidly than the mass [as indeed cosmic censorship suggests], so that, if the evaporation did continue indefinitely, the KMV$_0$ black hole could eventually be approximated by a non-rotating planar hole. These \emph{do} have a positive specific heat under all circumstances \cite{kn:surya}, contradicting the assumption of indefinite evaporation; so the claim is very reasonable. Thus we can assume that, as usual in applications of the AdS/CFT duality, a relatively slowly rotating KMV$_0$ black hole ---$\,$ recall that this is the only kind we need here ---$\,$ is dual to a thermal gauge theory on the boundary.

Henceforth, we regard a = J$^*$/M$^*$ and T as the basic physical variables of the black hole; think of them as ``known". This is natural if we wish to use duality, because J$^*$/M$^*$ and T have direct meanings in the dual field theory: the former as the ratio of the angular momentum and energy densities of the spinning QGP, the latter as its temperature. The geometric black hole parameters M$^*$ and r$_{\m{h}}$ can be derived from them: equation (\ref{FAUCET}) can always be solved for r$_{\m{h}}$, and then M$^*$ can be computed using (\ref{FLOPPY}). Thus r$_{\m{h}}$ and M$^*$ will be regarded as known functions of a and T, which can be prescribed independently of each other.

As in the case of rotating black holes with topologically spherical event horizons, the event horizon here ``rotates" as if it were a rigid object; that is, the angular velocity $\omega$(r$_{\m{h}}$) is a constant [equal to a/r$_{\m{h}}^2$], independent of $\psi$. For all other values of r, however ---$\,$ \emph{including} r $\rightarrow \infty$ ---$\,$ this is not so. One should therefore picture the spacetime as rotating differentially [or shearing], in a manner that depends both on distance from the black hole and on the azimuthal coordinate.

We now turn to the question of the geometry induced by the KMV$_0$ black hole at infinity. Note first that the angular velocity at infinity depends on the square of $\psi$:
\begin{equation}\label{eq:NUISANCE}
\m{\omega_{\infty}\;=\;a\psi^2/L^2.}
\end{equation}
Using this, one finds after a straightforward calculation that the geometry at infinity is described by the metric
\begin{equation}\label{NANNY}
\m{g(KMV_0)_{\infty}\; = \: -\,dt^2 \;-\;  2a\psi^2dtd\zeta \;+\; L^2\Big({d\psi^2\over 1 + a^2 \psi^4/L^2} +  d\zeta^2 \Big).}
\end{equation}
Setting d$\xi$ = d$\psi$/$\m{\sqrt{1 + a^2 \psi^4/L^2}}$, we see that this metric is identical to the one given in the preceding section by equation (\ref{C}), provided that we set $\tau$ = t/L, $\alpha^2$ = a/L, and that the function $\Upsilon(\xi \,,\alpha)$ takes a specific form given by inverting an elliptic integral: it is in fact given by
\begin{equation}\label{NONG}
\m{\Upsilon (\xi\,,\alpha) \; = \; \gamma^{- 1}sn\Big(\gamma \alpha \xi \; , \; i\Big),}
\end{equation}
where sn(x , k) denotes one of the Jacobi elliptic functions and $\gamma$ is an eighth root of unity. This can be further simplified to an explicitly real form if necessary by consulting Chapter 17 of \cite{kn:abramo}; one can show that it vanishes when either $\xi$ or a/L does so, and that it is indeed an increasing function of both $\xi$ and a/L, as is required if Figure 2 is to be replicated. Figure 3 shows the function $\Upsilon(\xi \,,1)$ corresponding to a KMV$_0$ metric with a/L = 1 [and suitably large M$^*$, so that the censorship condition (\ref{eq:N}) is satisfied], graphed on the horizontal axis for comparison with Figure 2.
\begin{figure}[!h]
\centering
\includegraphics[width=0.7\textwidth]{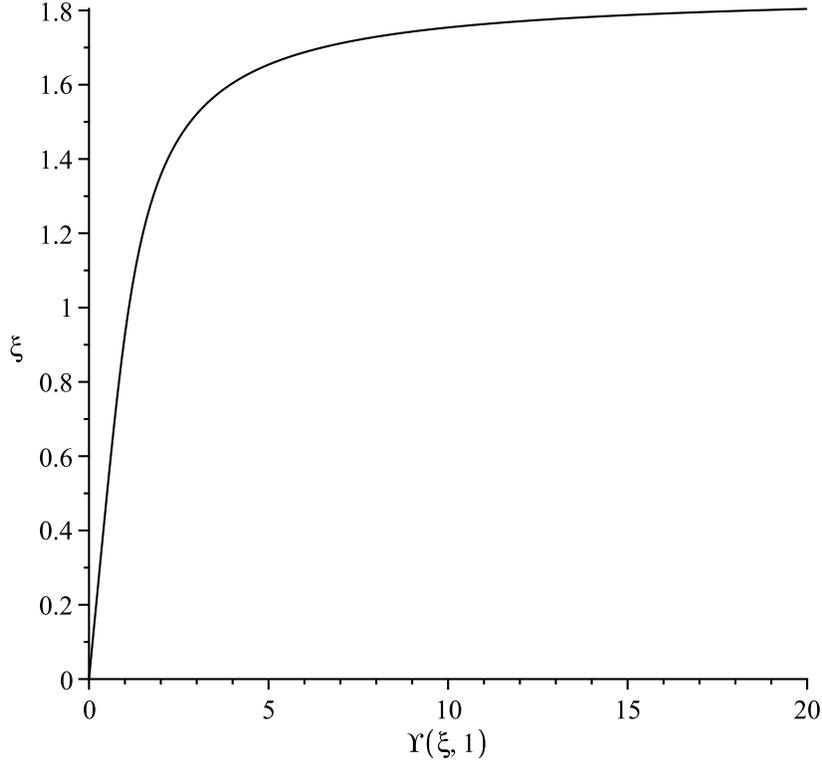}
\caption{$\Upsilon(\xi \,,1)$ for KMV$_0$ Black Hole [horizontal axis].}
\end{figure}

Recall from Section 2 that we only want values of x = L$\xi$ in the range 0 $\leq$ x $\leq$ X; that is, we only want $\xi$ in the range 0 $\leq$ $\xi$ $\leq$ X/L. [Similarly, we only want $\zeta$ in the range 0 $\leq$ $\zeta$ $\leq$ Z/L.] Since $\xi$ is a monotonically increasing function of $\psi$ and vice versa, this means that, in the KMV$_0$ black hole spacetime, we only need values of $\psi$ between zero and $\m{\sqrt{L\over a}\,\gamma^{- 1}sn\Big(\gamma \sqrt{a\over L}\,{X\over L}\; , \; i\Big)}$. This is why we stated above that $\psi$, like $\zeta$, only has a finite range. Actually, when $\Upsilon(\xi \,,\alpha)$ is given by equation (\ref{NONG}), it is only well-defined on a finite domain around zero, given [as in fact can be seen in Figure 3] approximately by $\m{- 1.85407\sqrt{L/a} < \xi < + 1.85407\sqrt{L/a}}$; so that $\xi$ has to be bounded for mathematical reasons. This means, since L and a are fixed by physical data, that one has to choose X to be sufficiently small [specifically, $\m{L^{3/2}\,> \,0.53935\,\sqrt{a}\,X}$], so that $\xi$ can range throughout 0 $\leq$ $\xi$ $\leq$ X/L. In practice, one would have to choose X to be small enough so that $\Upsilon(\xi \,,\alpha)$ does not become excessively large.

Thus we see that, for given values of the temperature and the angular momentum and energy densities, the KMV$_0$ black hole is in fact the dual of a field theory defined on a geometry of the sort constructed in Section 2, albeit for a particular [but not unreasonable] \emph{choice} of $\Upsilon(\xi \,,\alpha)$. We do not claim that this particular choice is physically realistic ---$\,$ that is, we do not claim that a realistic thickness function, combined with physical values for the temperature and angular momentum and energy densities in a spinning QGP, would lead to the expression in equation (\ref{NONG}), even for small values of X. One can however obtain other $\Upsilon(\xi \,,\alpha)$ functions by regarding the boundary metric as a boundary condition for the Einstein equations in the bulk, along the lines of \cite{kn:murata} [see also \cite{kn:kunz} and, for the general theory, \cite{kn:chrusc}]. This might require a sophisticated numerical investigation \cite{kn:vitor}. For our purposes, it will suffice to use the form given in equation (\ref{NONG}) [that is, to use the specific geometry of the KMV$_0$ black hole]; other $\Upsilon(\xi \,,\alpha)$ functions will be investigated elsewhere.

We are now in a position to investigate the question raised in Section 1: is this black hole, and consequently the dual system, unstable to the stringy effect discovered by Seiberg and Witten?

\addtocounter{section}{1}
\section* {\large{\textsf{4. The KMV$_0$ Black Hole in String Theory}}}
In \cite{kn:74} it was shown that AdS-Kerr black holes with topologically spherical event horizons are completely immune to the brane pair-production instability discussed in \cite{kn:seiberg}; surprisingly [because, in other examples, such severe deformations of the AdS-Schwarzschild geometry do generate an instability], this turned out to be true even for extremal black holes. One might expect that the relatively slowly rotating planar black holes discussed above should also be stable. Let us investigate.

In order to follow the procedure of Seiberg and Witten, we need to discuss the Euclidean version of the KMV$_0$ metric.
For each pair (a, M$^*$), the KMV$_0$ metric defines a Euclidean gravitational instanton, given by
\begin{equation}\label{O}
\m{g(EKMV_0) = {\Delta^E_r\Delta^E_{\psi}\rho_E^2\over \Sigma_E^2}\,dt^2\;+\;{\rho_E^2 \over \Delta^E_r}dr^2\;+\;{\rho_E^2 \over \Delta^E_{\psi}}d\psi^2 \;+\;{\Sigma_E^2 \over \rho_E^2}\Bigg[\omega_E\,dt \; - \;d\zeta\Bigg]^2},
\end{equation}
where, as is customary, we use the same names for Euclidean and Lorentzian coordinates, though Euclidean ``time" must, as always, be compactified. The coefficient functions are obtained by continuation of the angular momentum parameter, as usual:
\begin{eqnarray}\label{eq:P}
\m{\rho_E^2} & = & \m{r^2\;-\;a^2\psi^2} \nonumber\\
\m{\Delta^E_r} & = & \m{- a^2+ {r^4\over L^2} - 8\pi M^*r}\nonumber\\
\m{\Delta^E_{\psi}}& = & \m{1 - {a^2 \psi^4\over L^2}}\nonumber\\
\m{\Sigma_E^2} & = & \m{r^4\Delta^E_{\psi} + a^2\psi^4\Delta^E_r}\nonumber\\
\m{\omega_E} & = & \m{{\Delta^E_r\psi^2\,+\,r^2\Delta^E_{\psi}\over \Sigma_E^2}\,a}.
\end{eqnarray}
In order for this instanton to be well-defined, we clearly need $\m{\Delta^E_{\psi},\Sigma_E}$, and $\m{\rho_E}$ to be positive under all circumstances. For $\m{\Delta^E_{\psi}}$ to be positive, $\psi$ \emph{must} be confined to a finite range\footnote{That is, there is a topological identification in this direction in the Euclidean case [only]. Clearly $\psi$ is not a very satisfactory global coordinate on a circle, but, as explained in \cite{kn:klemm}, this can be rectified by setting up a new system of coordinates on overlapping coordinate charts.} $0 \, \leq \, \psi \, \leq \,\Psi$, where $\Psi$ has to satisfy
\begin{equation}\label{PP}
\m{\Psi\; < \; \sqrt{{L\over a}}},
\end{equation}
the inequality being strict. From the censorship inequality (\ref{eq:NOW}) in the preceding section, the natural choice for $\Psi$ is
\begin{equation}\label{eq:QQQQ}
\m{\Psi \; = \; {1 \over 3^{1/4}}\Bigg({L \over 2\pi M^*}\Bigg)^{1/3},}
\end{equation}
and we adopt this henceforth.

Next, the radial coordinate is only well-defined for r $\geq$ r$_0$, where r$_0$ satisfies
\begin{equation}\label{Q}
\m{\Delta^E_r(r_0)\;=\;0.}
\end{equation}
Since $\m{\Delta^E_r}$ increases beyond r = r$_0$, it is never negative, and so $\m{\Sigma_E}$ never vanishes if (\ref{PP}) holds.
Finally, the equation (\ref{Q}) can be manipulated into the form
\begin{equation}\label{QQ}
\m{\Bigg({r_0^2 \over a^2}\,-\,{L\over a}\Bigg)\Bigg({r_0^2 \over a^2}\,+\,{L\over a}\Bigg)\;=\;{8\pi M^*r_0L^2\over a^4},}
\end{equation}
whence it follows from (\ref{PP}) that
\begin{equation}\label{QQQ}
\m{{r^2\over a^2} \geq {r_0^2\over a^2} > {L\over a} > \Psi^2 > \psi^2}
\end{equation}
for a black hole of positive mass parameter; hence the condition that $\Delta_{\psi}$ should be positive [inequality (\ref{PP})] actually suffices to ensure that $\m{\rho_E}$ also never vanishes.

As long as we enforce (\ref{PP}), then, the metric has no pathologies apart, perhaps, from the familiar one arising from equation (\ref{Q}), which in general will give rise to a conical singularity. As in the topologically spherical case \cite{kn:hawrot}\cite{kn:solod}, avoiding this problem forces us to perform a further topological identification in the t-$\zeta$ plane: there is an identification under the mapping
\begin{equation}\label{QUASIMODO}
\m{(t,\;\zeta)\;\rightarrow \;(t\,+\,P,\;\zeta\,+\,Z_E )}.
\end{equation}
Here P becomes the reciprocal of the black hole temperature, 1/T, when the continuation to Lorentzian signature is performed, and Z$_{\m{E}}$ is given in terms of P by
\begin{equation}\label{R}
\m{Z_E = \omega_E (r_0) \times P \;= \; {\,a P \over r_0^2}}.
\end{equation}
Notice that the Euclidean version of the spacetime is necessarily of finite volume in the non-radial directions; in fact the transverse sections, and the boundary, have the topology of a three-dimensional torus with coordinates (t, $\psi$, $\zeta$). [This torus is not, however, a \emph{flat} torus in general.]

Note that (\ref{R}) is only valid for a $\neq$ 0; in the a = 0 case the periodicity is not constrained. That case cannot be obtained by taking the limit. It requires a separate treatment, which was given in \cite{kn:conspiracy}. One finds that the action describing BPS branes outside a non-rotating planar [or toral] AdS black hole is positive and monotonically increasing at all points outside the event horizon. [Figure 4 shows the graph [of a convenient positive multiple] of the action function in this case for a typical choice of parameters.] This means that, if a brane-antibrane pair is created in this spacetime, the action can always be made smaller by moving closer to the event horizon: there is no tendency for the branes to expand, and the system is stable. By contrast, the brane action for non-rotating AdS black holes with \emph{negatively} curved event horizons is unbounded below [see \cite{kn:conspiracy}], and so the system is unstable in the sense discussed by Seiberg and Witten \cite{kn:seiberg}.
\begin{figure}[!h]
\centering
\includegraphics[width=0.7\textwidth]{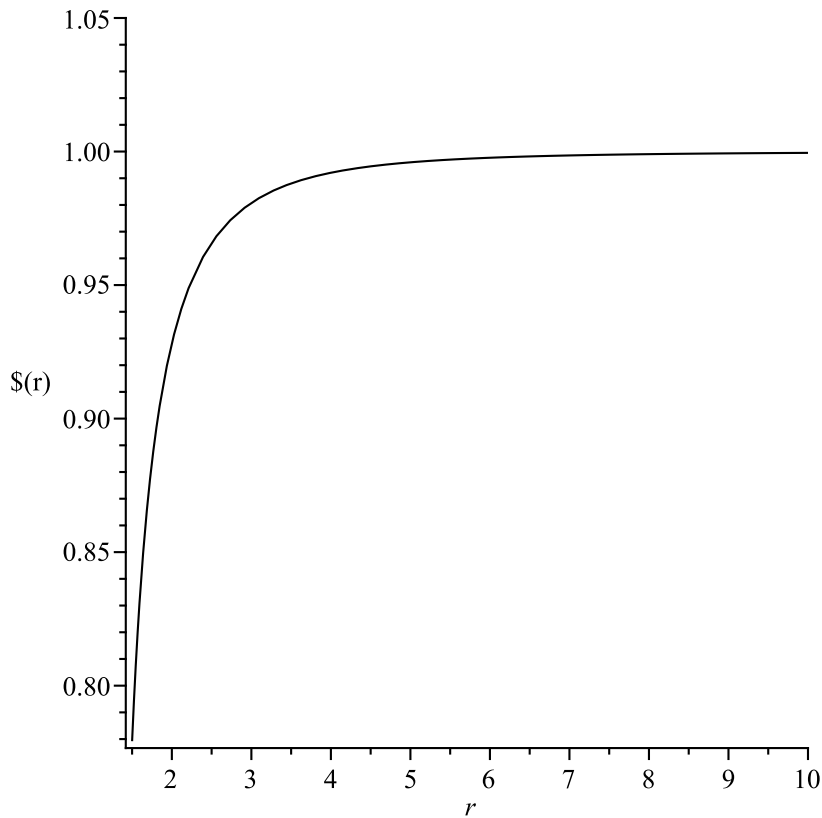}
\caption{Brane Action for a = 0, Typical Parameter Values}
\end{figure}

Returning to the a $\neq$ 0 case: the Euclidean action of a BPS 2-brane located at a given value of r comprises two competing contributions, one related to the area of the brane and the other to its volume \cite{kn:seiberg}. [Because all of the non-radial directions are, for one reason or another, of finite extent, these areas and volumes are all finite.] The total is given in this case by
\begin{equation}\label{eq:RR}
\m{E\$[KMV_0] \;=\;\Theta \, Z_E P \Bigg[\int_0^{\Psi}\sqrt{\Delta^E_r}\,\rho_E \, d\psi \; - \; {3 \over L} \int_{r_0}^r \int_0^{\Psi}\rho_E^2 \,d\psi \, dr}\Bigg],
\end{equation}
where $\Theta$ is the brane tension, and P and Z$_{\m{E}}$ are as in (\ref{QUASIMODO})(\ref{R}). Performing the integrals and the analytic continuation back to Lorentzian signature, we find at length that the action per unit brane tension for a $\neq$ 0 is
\begin{eqnarray}\label{S}
\m{\$[KMV_0]/\Theta} \; & = & \;\m{{a \over T^2r_{h}^2}\Bigg\{{r^2\over 2} \sqrt{a^2+ {r^4\over L^2} - 8\pi M^*r}\Bigg[{1 \over a}arcsinh\Big({a\Psi\over r}\Big) + {\Psi \over r}\sqrt{1+{a^2\Psi^2\over r^2}}\Bigg]} \nonumber \\
& & \;\;\;\;\;\;\;\;\;\;\;\;\;\;\;\;\;\;\;\; - \m{ {1\over L}\Bigg[\Psi(r^3 - r_{h}^3) + a^2\Psi^3(r-r_{h})\Bigg]\Bigg\},}
\end{eqnarray}
where we remind the reader that r$_{\m{h}}$ is to be computed from the basic variables T and a by using (\ref{FAUCET}), M$^*$ is found from (\ref{FLOPPY}), and $\Psi$ is given by equation (\ref{eq:QQQQ}). This equation is valid only at the event horizon [where the action is zero] and outside it.

Because of the term involving the square root of a quantity which vanishes at the event horizon, the graph of the action function has a large positive slope
near to the event horizon, and so the action must be positive in that region. When r is large, however, the leading terms contributed by both the area and the volume are proportional to r$^3$, this being typical of asymptotically AdS geometries. The coefficients of these terms are exactly equal, so the leading terms cancel; Seiberg and Witten explain that this is due to the BPS condition on the brane charge. At large values of r, then, the action is determined by two linear terms of opposite signs; these do not cancel, and in fact one finds that the term contributed by the volume is larger:
\begin{equation}\label{eq:T}
\m{\$[KMV_0]/\Theta \;=\; {a \over T^2r_{h}^2}\Bigg\{{ -\,5a^2\Psi^3 \over 6L}\,r \;+\;{\Psi r_{h} \over L}\Bigg[\,r_{h}^2 \;+\;a^2\Psi^2 \;-\;{4\pi M^*L^2 \over r_{h}}\Bigg] \;+\; O(1/r) \Bigg\}}.
\end{equation}
At large r, then, the action function behaves like a linear function with a negative slope, which means that the action will be \emph{negative} beyond some value of r, and in fact unbounded below.

\begin{figure}[!h]
\centering
\includegraphics[width=0.7\textwidth]{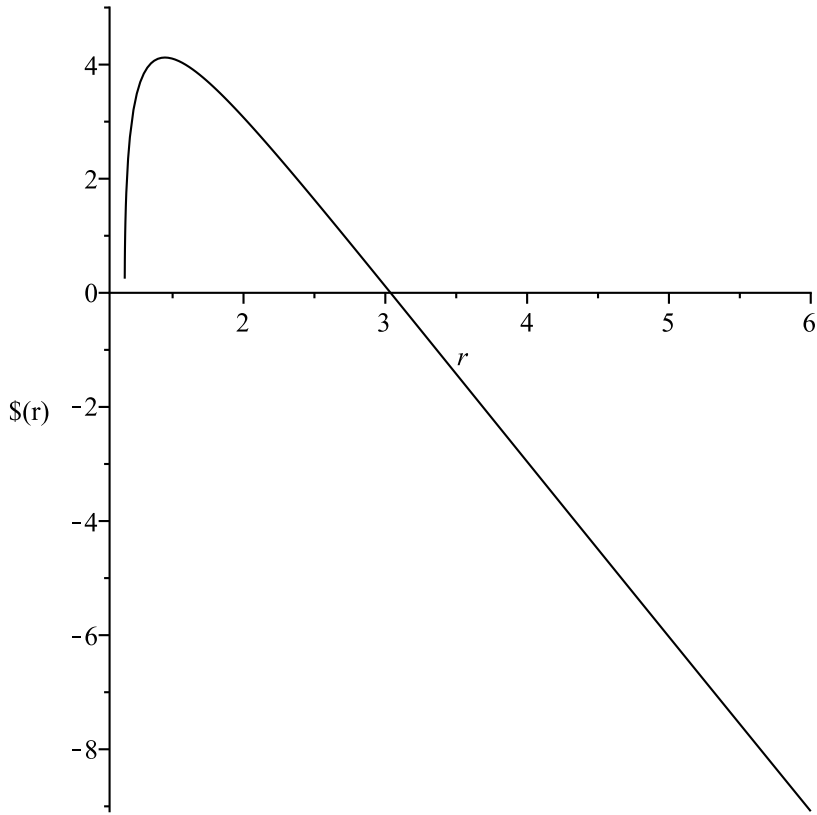}
\caption{KMV$_0$ Brane Action, Unit Tension,  a$^2$/L$^2$ = 1/2, 4$\pi$M$^*$/L = 1.}
\end{figure}

Let us examine a concrete example. Take 4$\pi$M$^*$/L = 1, a$^2$/L$^2$ = 1/2. Then the right side of (\ref{eq:N}) is approximately 1.1906, so cosmic censorship is satisfied by a wide margin; the event horizon is at r$_{\m{h}}$ $\approx$ 1.1622L, $\Psi \approx 0.9573$, and the temperature is $\approx$ 0.2521/L. Figure 5 shows the shape of the graph for a brane of unit tension; as expected, it rises rapidly at first, but then turns over and quickly comes to resemble a linear function with a negative slope.

This means that, if a brane is formed by pair-production sufficiently far from the event horizon, it will always find that it can lower its action locally [and, eventually, globally, that is, below the value at the event horizon] simply by expanding: the system is unstable. This is true for all non-zero values of a. Thus we find that \emph{any amount of rotation causes a topologically planar AdS black hole to become unstable}. This is very different from the case of branes nucleating in the spacetime around an AdS-Kerr-Newman black hole with a topologically \emph{spherical} event horizon: in that case, the action is always positive and monotonically increasing outside the event horizon, and so branes always tend to contract \cite{kn:74}, as in the a = 0 planar case discussed above. It is also contrary to expectations based on the AdS-Reissner-Nordstr$\ddot{\m{o}}$m \emph{planar} black hole. In that case one finds \cite{kn:AdSRN}\cite{kn:triple} that the addition of electric charge does not necessarily cause the black hole to become unstable: this only happens with the addition of \emph{large} amounts of charge, approximately 92$\%$ of the extremal charge in the four-dimensional case \cite{kn:73}. [The action function in this case rises to a maximum and then decreases, as it does here; but its slope is asymptotically vanishing, not negative, so the asymptotic value of the function itself can have either sign.] One might say that angular momentum has a more destabilizing effect than electric charge.

In short, AdS black holes with topologically planar event horizons are ``fragile" \cite{kn:fragile} [see also \cite{kn:74}, and, for possible implications in higher dimensions, \cite{kn:brihaye}], in the sense that a small deformation, in this case due to the addition of angular momentum, renders them unstable. In the Introduction, we pointed out that one of the drawbacks of using black holes with topologically spherical event horizons [in this application] is that positive scalar curvature tends to mask a Seiberg-Witten instability which might be present in the actual physical system being modeled. We now see that this is indeed precisely what happens: as soon as we pass to a more realistic geometric structure on the boundary, the system does indeed immediately become fragile. Let us consider the meaning of this result.

\addtocounter{section}{1}
\section* {\large{\textsf{5. Questions of Causality}}}
In terms of the diagrams, the effect of adding angular momentum to a planar asymptotically AdS black hole is readily understood: the graph in Figure 4 is asymptotically horizontal, but the effect of angular momentum is to cause it to bend downwards, producing a graph like the one in Figure 5. No matter how slight the bending, the resulting graph must eventually cut the horizontal axis, and the action will be negative from that point onwards.

Physically, this seems unreasonable: it means, as we have seen, that the system becomes unstable for any amount of angular momentum, no matter how small. According to the AdS/CFT correspondence, the same statement holds for the dual system, that is, for the QGP formed from peripheral heavy ion collisions. Since, in reality, every such collision deposits \emph{some} amount of angular momentum, however small, into the interaction zone, our result seems to indicate that the QGP, as formed in such collisions, is \emph{always} unstable. Even leaving aside the possible conflict with observations, this is not credible.

Intuition suggests that the instability, if indeed it exists, should not have any observable effect on the QGP when its angular momentum density is non-zero but small. Let us see how that might work.

The Seiberg-Witten instability can be studied in the case of non-rotating, asymptotically AdS black holes with event horizons having the geometry of a compact space of constant negative curvature. As was mentioned earlier, these objects are \emph{always} unstable \cite{kn:conspiracy}, for all values of the parameters, but it was shown in \cite{kn:barbon} that for temperatures which are high relative to 1/L, the reciprocal of the curvature scale, the \emph{rate} at which branes nucleate is suppressed. Black holes of this kind are very different from those studied here, particularly because they have no continuous parameter analogous to the angular momentum, but let us assume that there is a similar effect in our case; the graph of the brane action is in fact somewhat similar in the two cases [when the angular momentum in our case is large]. Now the QGP is the state of quark matter for arbitrarily high temperatures: but such temperatures [attained for example in the early Universe or possibly in the early stages of a supernova] do not persist indefinitely. This prompts the following suggestion: perhaps the QGP is indeed always unstable to brane pair-production, but \emph{the rate at which the pairs nucleate is, at very extreme temperatures, so low that no such effect can be observed on the relevant time scales}.

Conversely, however, there is no such suppression of the rate when T $\approx$ 1/L. This regime has the following physical interpretation. The familiar non-spinning QGP defines a fundamental temperature scale, given by the \emph{critical temperature}, below which the QGP gradually ceases to be the appropriate description of the system. [That is, there is a crossover rather than an abrupt phase change, at the low values of the chemical potential relevant to this discussion.] Current estimates of this temperature are T$_{\m{c}}$ $\approx$ 175 MeV; see for example \cite{kn:behan} and \cite{kn:ohnishi}. On the bulk side of the correspondence, the theory we have been using [small perturbations around Einstein gravity for asymptotically AdS spacetimes] is a good approximation to the  full string theory precisely when the curvature scale L is large relative to the fundamental string scale. The natural choice for L here is the scale defined by 1/T$_{\m{c}} \approx$ 7 femtometres, which is indeed ``large" in this sense.

With this understanding, we see that the rate at which branes nucleate in the bulk dual to a spinning QGP will \emph{not} be suppressed by an effect like the one discussed in \cite{kn:barbon} when the temperature of the plasma is not much above the critical temperature T$_{\m{c}}\,\approx \,$ 1/L. Our task now is reduced to explaining why no such effect is observed in this temperature regime \emph{at relatively low angular momenta}.

We can begin to understand this in the following way. It is clear from Figure 5 that, even for values of r which are not very much larger than r$_{\m{h}}$, the action per unit tension function is well approximated by its linearization. It is elementary to compute where this linearization [equation (\ref{eq:T})] cuts the horizontal axis, and so we obtain a good approximation for the value of r where the action becomes negative, r = r$_{\m{neg}}$ :
\begin{equation}\label{U}
\m{r_{neg} \; \approx \; {3\sqrt{3}\over 5 \times (4)^{2/3}}\,\Bigg({r_h^3\over a^2} \;- \; {L^2\over r_h}\Bigg)\Bigg({r_h^3\over L^3}\;+\;{a^2\over r_h L}\Bigg)^{2/3}},
\end{equation}
where equations (\ref{FLOPPY}) and (\ref{eq:QQQQ}) have been used; here we regard T as fixed. One finds that r$_{\m{neg}}$ becomes large as the angular momentum is reduced; see for example Figure 6.
\begin{figure}[!h]
\centering
\includegraphics[width=0.7\textwidth]{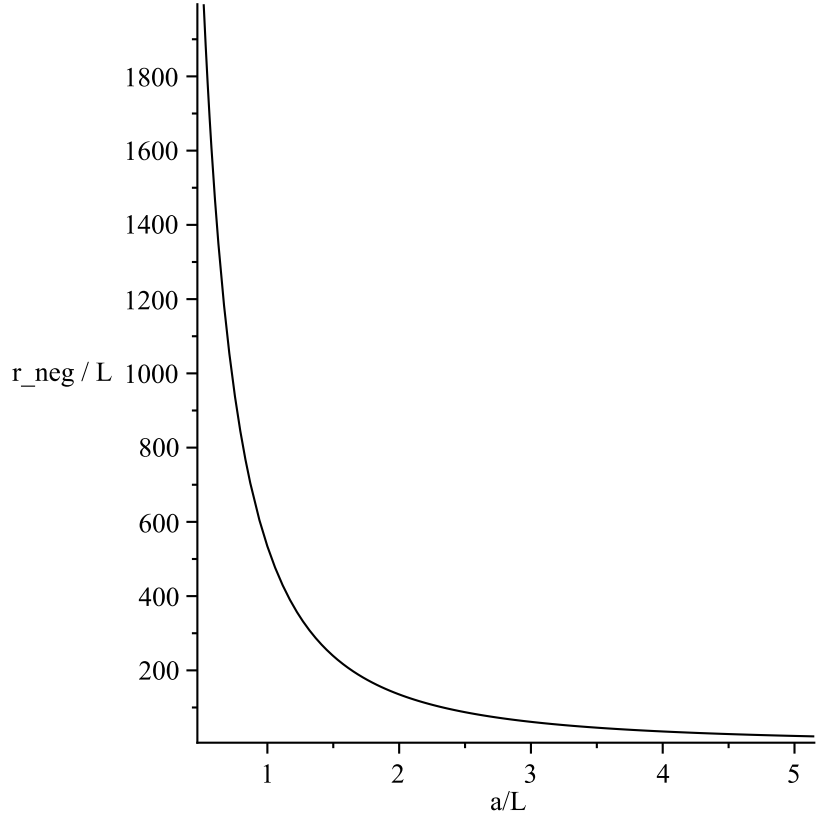}
\caption{r$_{\m{neg}}$/L as a Function of a/L, with T = 1/L.}
\end{figure}
Notice\footnote{Figure 6 is drawn by regarding r$_{\m{neg}}$ as a function of a/L through a parametric interpretation of equations (\ref{FAUCET}) and (\ref{U}), with T fixed at 1/L, and with r$_{\m{h}}$/L as the parameter. Here r$_{\m{h}}$/L varies between 4.19 and 4.30.} that r$_{\m{neg}}$/L is already very large when a/L $\approx$ 1; by the time a/L falls to 0.5, r$_{\m{neg}}$/L is already around 2000. On the other hand, if we fix T and regard M$^*$ and r$_{\m{h}}$ as functions of a [through equations (\ref{FAUCET}) and (\ref{FLOPPY})], then we find that r$_{\m{h}}$ is a monotonically increasing function of a, so it becomes smaller [but approaches a non-zero limit] as a is reduced.

In short, r$_{\m{neg}}$ $-$ r$_{\m{h}}$ increases indefinitely as the angular momentum is reduced. That is: if we take a non-rotating planar black hole [which is stable] and give it a very small amount of angular momentum, then the brane action will indeed become negative; but the region of negative action moves \emph{inward} ``from infinity". In other words, for small angular momenta, it is far away from the event horizon. [The coordinate r is admittedly not a good measure of proper distance, but the latter does diverge when r diverges.]

Simple classical causality suggests that this means that the instability \emph{of the whole bulk spacetime} develops only slowly when the angular momentum of the hole is small: intuitively, it takes time for the misbehaviour of the brane action far from the black hole to make itself felt in the vicinity of the event horizon. But the non-local nature of the AdS/CFT correspondence means that this is indeed the relevant time scale: the boundary theory is dual to the entire bulk, not just the special region in which the branes begin to misbehave. In other words, we propose that causality in the KMV$_0$ black hole geometry should give rise to a time scale which, in the dual interpretation, computes the time scale for the instability to make itself felt in the spinning QGP [in those cases where the rate of brane nucleation is not already suppressed by extremely high temperatures, as above]. The upshot is that small angular momenta \emph{delay} the development of the instability, just as high temperatures suppress the nucleation rate.

Now however we have a simple but crucial observation: the QGP, as formed in a heavy ion collision, immediately begins to expand and cool. Within a very short time it will hadronize. \emph{Any effect which takes longer than this to evolve will not be observed} ---$\,$ the QGP no longer exists. We suggest that this explains why Seiberg-Witten instability is not observed at low angular momenta: in such a plasma, the instability is delayed too long to be observed. Conversely, at high angular momenta the delaying effect is removed or at least reduced, and one might expect the effect to become observable.

The conclusion, then, is that the Seiberg-Witten instability of the KMV$_0$ spacetime is likely to have observable effects in the QGP when the latter has the following two properties:

[a] Its temperature is not far above the critical temperature [$\approx$ 175 MeV], so that the rate of nucleation of branes in the dual geometry is not suppressed by the effect described in \cite{kn:barbon}. That is, we are considering the QGP in its final stages, shortly before it hadronizes.

[b] Its angular momentum density is so high that the region in which the branes nucleate in the dual geometry is sufficiently near to the event horizon that there is no significant delay in propagating the effects arising in the negative-action region to the entire bulk.

How might the delay time in [b], corresponding to a given temperature and angular momentum parameter, be estimated? A key observation here \cite{kn:gibbo} is that the AdS curvature scale L actually finds its main physical interpretation \emph{as a time scale}: every free massive particle emitted from a given point in space returns to that point after a proper time $\pi$L [which is independent of the initial velocity]. To put it in a different way, more suited to the generalization to asymptotically AdS spaces: the proper time required for a massive object to fall from rest, relative to a Killing observer, to the origin is $\pi$L/2. For AdS itself, the time required is actually independent of the initial distance from the origin, as is the case for any simple harmonic oscillator.

The analogous quantities for an asymptotically AdS black hole spacetime are the proper times required for an object released by a suitably distinguished observer to reach the event horizon. For the case at hand, we are interested in an observer located in the KMV$_0$ spacetime at the point where the brane action becomes negative: we wish to compute the proper time required for an object to fall from r = r$_{\m{neg}}$ to r = r$_{\m{h}}$. We found above that r$_{\m{neg}}$ $-$ r$_{\m{h}}$ depends strongly on the angular momentum. In AdS itself, as we have just seen, this would have no effect on the proper time of fall, but, in a less symmetric spacetime which is merely \emph{asymptotically} AdS, it does: the distortion of the spacetime due to rotation will cause the time of fall to vary with the initial position.

Before proceeding to compute this time, we should discuss the observational situation, without pretending to be very precise. Using data from the RHIC experiment, Becattini et al. \cite{kn:bec} estimate that the maximal amount of angular momentum transferred to the plasma in that case is around 72000 [angular momentum being dimensionless in natural units]. This occurs at the optimal impact parameter of about 2.5 fm and at a temperature of perhaps twice the critical temperature. For \emph{central} collisions, the interaction region is estimated in \cite{kn:phobos} to have a transverse area of 150 fm$^2$ and an [effective] longitudinal size at equilibrium of roughly 2 fm. Using the classical formula for the area of overlapping circles \cite{kn:wolfram}, we can compute the volume of the interaction region for an optimal peripheral collision, and so obtain an approximate angular momentum density of around 360/fm$^3$. The energy density is roughly \cite{kn:phobos} 3 - 5 GeV/fm$^3$, so the ratio of the densities is around 100 - 150 fm. At the LHC, the corresponding figure could ultimately be well over an order of magnitude larger \cite{kn:bec}. Thus we should \emph{not} assume that our parameter a, which measures the ratio of the angular momentum and energy densities of the black hole, need be of order unity; it ranges from near zero, for very central collisions, up to over a hundred fm at the RHIC. If we take T$_{\m{c}}$ $\approx$ 175 MeV as above, then L $\approx$ 7 fm. We conclude that a/L, too, need not be small; it ranges from around zero up to perhaps $\approx$ 20 at the RHIC, and possibly 1000 at the LHC. These figures are very approximate; we present them in order to clarify what we mean by ``small" and ``large" angular momenta, and to stress that, in the actual experiments, the dimensionless parameter a/L takes its values in a wide range.

Let us return to the proper time of fall. Call it s(a/L, TL). It depends non-trivially on the basic dimensionless parameters a/L and TL; granted this, we expect it to be a decreasing function of a/L, since the region of negative action moves closer to the event horizon as a/L increases. Now an important point is this: we do not want s(a/L, TL) to be a rapidly varying function of a/L. For we saw above that a/L has a very wide range in the actual experiments, from around zero up to perhaps a thousand. If s(a/L, TL) varies rapidly with a/L, then the delaying effect of low angular momenta will quickly be eliminated as the angular momentum increases, and the instability of the plasma will make itself felt in nearly central collisions at relatively low energies, contrary to observations. There is no reason a priori why s(a/L, TL) should vary exceptionally slowly with a/L, and so we have a non-trivial check of the general plausibility of our proposal. [On the other hand, of course, we do not want it to vary so slowly that the effect is completely unobservable.]

The function s(a/L, TL) can readily be computed if we confine ourselves to the $\psi$ = 0 plane, parametrized by r and $\zeta$. [From the first equation in the set (\ref{eq:M}), one sees that $\rho$ $>$ r for any other value of $\psi$, so on the curve parameterized by $\psi$ and defined by r = constant, $\zeta$ = constant, the point with $\psi$ = 0 is distinguished by being the one with the smallest proper distance to the event horizon.] Suppose that the object is released at rest by an observer with zero angular momentum. The equations of motion, derived in the usual way from the existence of the Killing vector fields $\partial_{\m{t}}$ and $\partial_{\zeta}$, are
\begin{equation}\label{eq:X}
\m{\dot{t}\Bigg[{-\,\Delta_r \over r^2}\;+\;{a^2 \over r^2}\Bigg]\;-\;\dot{\zeta}a\;=\;constant,}
\end{equation}
\begin{equation}\label{eq:Y}
\m{-\,\dot{t}a\;+\;\dot{\zeta}r^2\;=\;0},
\end{equation}
where the dot denotes the derivative with respect to proper time; this is to be supplemented with the condition that the worldline be timelike:
\begin{equation}\label{eq:Z}
\m{\dot{t}^2\Bigg[{-\Delta_r \over r^2}\;+\;{a^2\over r^2}\Bigg]\;-\;2a\dot{t}\dot{\zeta}\;+\;\dot{\zeta}^2r^2\;+\;{\dot{r}^2r^2\over \Delta_r}\;=\;-\,1.}
\end{equation}
Combining these equations and using the initial condition [that the derivative of r should vanish at r = r$_{\m{neg}}$] we obtain
\begin{equation}\label{ALPHA}
\m{\dot{r}^2\;=\;{\Delta_r(r_{neg})\over r_{neg}^2}\;-\;{\Delta_r\over r^2}.}
\end{equation}
Integrating this between r = r$_{\m{neg}}$ and r = r$_{\m{h}}$, we obtain the desired proper time:
\begin{equation}\label{BETA}
\m{{s(a/L, TL)\over L}\;=\;\int_{r_{h}}^{r_{neg}} \Bigg[{\Delta_r(r_{neg})\over r_{neg}^2}\,-\,{a^2\over r^2}\,-\,{r^2 \over L^2}\,+\,{8\pi M^*\over r}\Bigg]^{-\,1/2}{dr\over L}.}
\end{equation}
Here we regard TL as fixed; then once a/L is specified, r$_{\m{h}}$/L is computed from equation (\ref{FAUCET}), then M$^*$/L from (\ref{FLOPPY}), then $\Psi$ from equation (\ref{eq:QQQQ}), then r$_{\m{neg}}$/L by solving the equation obtained by setting the right side of equation (\ref{S}) equal to zero, and so finally s(a/L, TL)/L can be computed from equation (\ref{BETA}).

Any attempt to draw quantitative conclusions from this simple model is of course to be treated with due scepticism. Nevertheless, the results are suggestive and will help us to explain the qualitative conclusions which will be discussed below. Let us suppose that we have a plasma at a temperature of T $\approx$ 2 $\times$ T$_{\m{c}}$; it is cooling rapidly towards the critical temperature. We have numerically evaluated s(a/L, TL)/L at T = 2 $\times$ T$_{\m{c}}$, over the range of values of a/L discussed above. [We have also done this for T = T$_{\m{c}}$, so that the variation of s(a/L, TL)/L can be seen as the temperature drops towards the critical temperature.]
\begin{center}
\begin{tabular}{|c|c|c|}
  \hline
a/L & s(a/L, TL = 1)/L  &  s(a/L, TL = 2)/L \\
\hline
1 &    1.56302    &  1.57006   \\
5 &   1.41813  &  1.55873 \\
100 &    0.63686  &  0.87705 \\
1000 &    0.51255  &  0.56674  \\
5000 &     0.48637            &    0.50818            \\
\hline
\end{tabular}
\end{center}
We see that s(a/L, TL)/L \emph{varies extremely slowly} with a/L: as a/L increases from near zero to 5000, it drops only to around one third of its initial value. As we discussed earlier, this is exactly what we need. [Notice that it also varies quite slowly with respect to temperature, especially at very high [and also very low] angular momenta. Thus one can argue that the cooling itself probably does not have much effect on these calculations.] It also confers the benefit that it renders our quantitative discussions somewhat more robust.

Our suggestion is that when [for example] a/L = 5, so that the instability time scale at TL = 2 is 1.55873 $\times$ L $\approx$ 11 fm/c, one may find that this is longer than the time required for the temperature of the plasma to drop from $\approx$ 2 $\times$ T$_{\m{c}}$ to below the critical temperature, so there is no time for the instability to make itself felt. By contrast, at a/L = 1000, which may possibly be attainable at the LHC, the scale is 0.56674 $\times$ L $\approx$ 4 fm/c, and this could be short enough to affect the plasma before it hadronizes. While these numbers are not to be taken at face value, they do suggest that the effect is not entirely out of reach.

In principle, then, we may have an explanation of the non-observation of Seiberg-Witten instability in nearly central heavy ion collisions. Equally, the theory indicates that the QGP may be unstable if it spins too rapidly. In practice this would mean that collisions involving very high angular momenta should simply fail to produce a plasma: \emph{there may be a ``holographic" angular momentum cutoff} in heavy ion collisions.

\addtocounter{section}{1}
\section* {\large{\textsf{6. Conclusion}}}
We have argued that the AdS/CFT dual of the QGP produced in a peripheral heavy ion collision is a particular kind of black hole characterized by its temperature T and the ratio, a, of its angular momentum and energy densities. This black hole is unstable in string theory when a is not zero, indicating that the spinning QGP is also unstable. We have argued, however, that low angular momenta delay the evolution of this effect, so that it cannot appear in the brief time required for the QGP to hadronize. On the other hand, there is a possibility that the instability will appear at sufficiently high angular momenta.

It is customary to warn \cite{kn:mateos}\cite{kn:karch} that the AdS/CFT correspondence must be applied to actual field theories with considerable circumspection, and this certainly applies here. At this point, then, one can only make a qualitative suggestion: that it may be worthwhile to look for evidence of a cutoff in the observable angular momenta of the QGP at very high values. [``High" here would mean that the ratio of angular momentum to energy densities might be large in units defined by the critical temperature.]

Unfortunately, it appears to be very difficult to find such evidence, even at this qualitative level. It is difficult in the first place to detect the quark polarization effects which might allow an exploration of the spinning QGP itself; see \cite{kn:huang} for a brief discussion and references. Notice too that there may be a hint of some form of peripheral-collision instability in the observed relationships [see for example \cite{kn:heinz}] between centrality and the elliptic flow observable v$_2$, which indicate that the hydrodynamic approximation begins to break down for these collisions. Overall, the prospects for identifying an angular momentum cutoff in heavy-ion collisions do not seem very promising at present. A more realistic approach may be to extend the ideas of this work to analogous phenomena in condensed matter theory [see for example \cite{kn:mcgreevy}], and to try to see whether the analogue of the angular momentum cutoff might have consequences in that case ---$\,$ for example, in the systems studied in \cite{kn:sonner}.

One can certainly work towards more sophisticated and convincing versions of the model put forward here. For example, one would want to study the possible effects of a dynamic spacetime geometry [arising from the fact that the temperature of the QGP, and therefore of the black hole, is far from constant], using the theory described in \cite{kn:veronika}, so as to confirm our suggestion that the cooling itself does not strongly affect the time scale of the instability. The effects of incorporating a non-zero chemical potential for the field theory would also be of interest; the chemical potential in the current experiments is not large, but this will change at, for example, FAIR \cite{kn:fair}.

A more elementary [but equally important] question, however, is this. We have arrived at our main conclusion, that the spinning QGP is in some way unstable, with the aid of a specific $\Upsilon(\xi \,,\alpha)$ function in equation (\ref{C}), the one presented to us by the KMV$_0$ geometry. While the latter is distinguished by its degree of symmetry and simplicity, other choices are possible and might be better motivated, for example by starting with a more realistic thickness function for the nuclei. Although it seems implausible that any such choice could drastically alter the shape of the graph in Figure 5, to the extent of removing the region of negative action altogether, one would like to settle this question definitively. We claim in fact that the instability will certainly arise for \emph{any} $\Upsilon(\xi \,,\alpha)$ function which is even roughly realistic, so that our principal conclusions cannot be avoided in this way. This will be discussed elsewhere.

\addtocounter{section}{1}
\section*{\large{\textsf{Acknowledgements}}}
The author is very grateful to everyone at the CECs, where this work was done, for making his visit such an enjoyable and fruitful one. He is also grateful to Prof. Soon Wanmei for support and encouragement and for help with the diagrams, and to Jude McInnes for assistance with translations.

\end{document}